\documentclass[aps,prd,twocolumns,eqsecnum,preprintnumbers,showpacs,amsmath,amssymb]
{revtex4}

\usepackage{graphicx}
\usepackage{dcolumn}
\usepackage{bm}

\begin{document}

\title{THE COLOR GAUGE INVARIANCE AND A POSSIBLE ORIGIN OF THE
JAFFE-WITTEN MASS GAP IN QCD}

\author{V. Gogokhia}
\email[]{gogohia@rmki.kfki.hu}

\affiliation{HAS, CRIP, RMKI, Depart. Theor. Phys., Budapest 114,
P.O.B. 49, H-1525, Hungary}

\date{\today}
\begin{abstract}
The physical meaning of a mass gap introduced by Jaffe and Witten
is to be responsible for the large-scale (low-energy/momentum),
i.e., the non-perturbative structure of the true QCD vacuum. In
order to make the existence of a mass gap perfectly clear it is
defined as the difference between the regularized full gluon
self-energy and its subtracted (also regularized) counterpart. The
mass gap is mainly generated by the nonlinear interaction of
massless gluon modes. A self-consistent violation of $SU(3)$ color
gauge invariance/symmetry is discussed in order to realize a mass
gap in QCD. For this purpose, we propose not to impose the
transversality condition on the full gluon self-energy, while
restoring the transversality of the full gluon propagator relevant
for the non-perturbative QCD at the final stage. At the same time,
the Slavnov-Taylor identity for the full gluon propagator is
always preserved. All this allows one to establish the general
structure of the full gluon propagator in the presence of a mass
gap. In this case, two independent types of formal solutions for
the full gluon propagator have been established. The nonlinear
iteration solution at which the gluons remain massless is
explicitly present. The existence of the solution with an
effective gluon mass is also demonstrated.
\end{abstract}



\maketitle

\section{Introduction}

Quantum Chromodynamics (QCD) \cite{1,2} is widely accepted as a
realistic quantum field gauge theory of strong interactions not
only at the fundamental (microscopic) quark-gluon level but at the
hadronic (macroscopic) level as well. This means that in principle
it should describe the properties of experimentally observed
hadrons in terms of experimentally never seen quarks and gluons,
i.e., to describe the hadronic world from first principles -- an
ultimate goal of any fundamental theory. But this is a formidable
task because of the color confinement phenomenon, the dynamical
mechanism of which is not yet understood, and therefore the
confinement problem remains unsolved up to the present days. It
prevents colored quarks and gluons to be experimentally detected
as physical ("in" and "out" asymptotic) states which are colorless
(i.e., color-singlets), by definition, so color confinement is
permanent and absolute \cite{1}.

Today there is no doubt left that color confinement and other
dynamical effects, such as spontaneous breakdown of chiral
symmetry, bound-state problems, etc., being essentially
nonperturbative (NP) effects, are closely related to the
large-scale (low-energy/momentum) structure of the true QCD ground
state and vice-versa (\cite{3,4} and references therein). The
perturbation theory (PT) methods in general fail to investigate
them. If QCD itself is a confining theory then a characteristic
scale has to exist. It should be directly responsible for the
above-mentioned structure of the true QCD vacuum in the same way
as $\Lambda_{QCD}$ is responsible for the nontrivial perturbative
dynamics there (scale violation, asymptotic freedom (AF)
\cite{1}).

However, the Lagrangian of QCD \cite{1,2} does not contain
explicitly any of the mass scale parameters which could have a
physical meaning even after the corresponding renormalization
program is performed. The main goal of this paper is to show how
the characteristic scale (the mass gap, for simplicity)
responsible for the NP dynamics may explicitly appear in QCD. This
becomes an imperative especially after Jaffe and Witten have
formulated their theorem "Yang-Mills Existence And Mass Gap"
\cite{5}. In order to make the existence of a mass gap perfectly
clear it is defined as the difference between the regularized full
gluon self-energy and its subtracted (also regularized)
counterpart. The mass gap is mainly generated by the nonlinear
(NL) interaction of massless gluon modes. Our proposal to realize
a mass gap in QCD is not to impose the transversality condition on
the full gluon self-energy, while restoring the transversality of
the full gluon propagator relevant for NP QCD at the final stage.
At the same time, the Slavnov-Taylor (ST) identity for the full
gluon propagator (important for the renormalization) is always
preserved within our approach.

 As mentioned above, there is no place for the mass gap in
the QCD Lagrangian, so the only place when the mass gap may appear
is the corresponding system of dynamical equations of motion, the
so-called Schwinger-Dyson (SD) equations, which should be
complemented by the ST identities (\cite{1} and references
therein). The propagation of gluons is one of the main dynamical
effects in the true QCD vacuum. It is described by the
above-mentioned corresponding SD quantum equation of motion for
the full gluon propagator. The importance of this equation is due
to the fact that its solutions reflect the quantum-dynamical
structure of the true QCD ground state. The color gauge structure
of this equation is also one of the main subjects of our
investigation in order to find a way how to realize a mass gap in
QCD.

In the presence of a mass gap two different and independent types
of formal solutions for the regularized full gluon propagator have
been established. The general nonlinear iteration solution is
always severely singular at small gluon momentum, i.e., the gluons
remain massless, and this does not depend on the gauge choice. The
massive solution leads to an effective gluon mass, which depends
on the gauge choice. No truncations/approximations/assumptions and
no special gauge choice are made for the skeleton loop integrals
(contributing to the full gluon self-energy) in order to show the
existence of these solutions. Both solutions are valid in the
whole energy/momentum range, and thus they should preserve AF when
the gluon momentum goes to infinity.

\section{QED}

It is instructive to begin with a brief explanation why a mass gap
does not occur in quantum electrodynamics (QED). The photon SD
equation can be symbolically written down as follows:

\begin{equation}
D(q) = D^0(q) + D^0(q) \Pi(q) D(q),
\end{equation}
where we omit, for convenience, the dependence on the Dirac
indices, and $D^0(q)$ is the free photon propagator. $\Pi(q)$
describes the electron skeleton loop contribution to the photon
self-energy (the so-called vacuum polarization tensor).
Analytically it looks

\begin{equation}
\Pi(q) \equiv \Pi_{\mu\nu}(q) = - g^2 \int {i d^4 p \over (2
\pi)^4} Tr [\gamma_{\mu} S(p-q) \Gamma_{\nu}(p-q, q)S(p)],
\end{equation}
where $S(p)$ and $\Gamma_{\mu}(p-q,q)$ represent the full electron
propagator and the full electron-photon vertex, respectively. Here
and everywhere below the signature is Euclidean, since it implies
$q_i \rightarrow 0$ when $q^2 \rightarrow 0$ and vice-versa. This
tensor has the dimensions of a mass squared, and it is
quadratically divergent at least in the PT. To make the formal
existence of a mass gap perfectly clear, let us now, for
simplicity, subtract its value at zero. One obtains

\begin{equation}
\Pi^s(q) \equiv \Pi^s_{\mu\nu}(q) = \Pi_{\mu\nu}(q) -
\Pi_{\mu\nu}(0) = \Pi_{\mu\nu}(q) - \delta_{\mu\nu}\Delta^2
(\lambda).
\end{equation}
The explicit dependence on the dimensionless ultraviolet (UV)
regulating parameter $\lambda$ has been introduced into the mass
gap $\Delta^2(\lambda)$, given by the integral (2.2) at $q^2=0$,
in order to assign a mathematical meaning to it. In this
connection a few remarks are in order in advance. The dependence
on $\lambda$ (when it is not shown explicitly) is assumed in all
divergent integrals here and below in the case of the gluon
self-energy as well (see next section). This means that all the
expressions are regularized (including photon/gluon propagator),
and we can operate with them as with finite quantities. $\lambda$
should be removed on the final stage only after performing the
corresponding renormalization program. So through this paper the
mass gap is a "bare" one, i.e., it is only regularized. Whether
the regulating parameter $\lambda$ has been introduced in a
gauge-invariant way (though this always can be achieved) or not,
and how it should be removed is not important for the problem if a
mass gap can be "released/liberated" from the corresponding
vacuum. We will show in the most general way (not using the PT and
not choosing any special gauge) that this is impossible in QED and
might be possible in QCD.

The decomposition of the subtracted vacuum polarization tensor
(2.3) into the independent tensor structures is

\begin{equation}
\Pi^s_{\mu\nu}(q) = T_{\mu\nu}(q) q^2 \Pi^s(q^2) + q_{\mu}
q_{\nu}(q) \tilde{\Pi}^s(q^2),
\end{equation}
where both invariant functions $\Pi^s(q^2)$ and
$\tilde{\Pi}^s(q^2)$ are, by definition, dimensionless and regular
at small $q^2$, since $\Pi^s_{\mu\nu}(0) =0$ identically due to
the subtraction (2.3); otherwise they remain arbitrary. From this
relation it follows that $\Pi^s_{\mu\nu}(q) = O(q^2)$, i.e., it is
always of the order $q^2$. In this connection a few remarks are in
order. The subtraction (2.3) at zero point in QED is justified,
since it is abelian gauge theory, and therefore there is no the
self-interaction of massless photons, which can be source of the
singularities in the $q^2 \rightarrow 0$ limit. The vacuum
polarization tensor (2.2) has no infrared (IR) singularities in
this limit, at least in the PT. So in what follows we will
consider the above-mentioned invariant functions as regular at
small $q^2$, indeed. Also, here and everywhere below

\begin{equation}
T_{\mu\nu}(q)=\delta_{\mu\nu}-q_{\mu} q_{\nu} / q^2 =
\delta_{\mu\nu } - L_{\mu\nu}(q).
\end{equation}

In the same way, the photon self-energy (2.2) in terms of the
independent tensor structures is

\begin{equation}
\Pi_{\mu\nu}(q) = T_{\mu\nu}(q) q^2 \Pi(q^2) + q_{\mu} q_{\nu}
\tilde{\Pi}(q^2),
\end{equation}
where both invariant functions $\Pi(q^2)$ and $\tilde{\Pi}(q^2)$
are dimensionless functions; otherwise they remain arbitrary. Due
to the transversality of the photon self-energy

\begin{equation}
q_{\mu} \Pi_{\mu\nu}(q) =q_{\nu} \Pi_{\mu\nu}(q) =0,
\end{equation}
which comes from the current conservation condition in QED, one
has $\tilde{\Pi}(q^2)=0$, i.e., it should be the purely
transversal

\begin{equation}
\Pi_{\mu\nu}(q) = T_{\mu\nu}(q) q^2 \Pi(q^2).
\end{equation}
On the other hand, from the subtraction (2.3), on account of the
relation (2.4), and the transversality condition (2.7) it follows
that

\begin{equation}
\tilde{\Pi}^s(q^2)= - {\Delta^2(\lambda) \over q^2},
\end{equation}
which, however, is impossible since $\tilde{\Pi}^s(q^2)$ is a
regular function of $q^2$, by definition. So the mass gap should
be discarded, i.e., put formally to zero and, consequently,
$\tilde{\Pi}^s(q^2)$ as well, i.e.,

\begin{equation}
\Delta^2(\lambda)=0, \quad \tilde{\Pi}^s(q^2)=0.
\end{equation}
Thus the subtracted photon self-energy is also transversal, i.e.,
it satisfies the transversality condition

\begin{equation}
q_{\mu} \Pi_{\mu\nu}(q) =q_{\mu} \Pi^s_{\mu\nu}(q) =0,
\end{equation}
and coincides with the photon self-energy (see Eq. (2.3) at zero
mass gap). Moreover, this means that the photon self-energy does
not have a pole in the $q^2 \rightarrow 0$ limit in its invariant
function $\Pi(q^2)= \Pi^s(q^2)$. As mentioned above, in obtaining
these results neither the PT has been used nor a special gauge has
been chosen. So there is no place for quadratically divergent
constants in QED, while logarithmic divergence still can be
present in the invariant function $\Pi(q^2) = \Pi^s(q^2)$. It is
to be included into the electric charge through the corresponding
renormalization program (for these detailed gauge-invariant
derivations explicitly done in lower order of the PT see Refs.
\cite{2,6,7,8,9}).

Taking into account the subtraction (2.3), on account of the
relations (2.10), the photon SD equation (2.1) becomes equivalent
to

\begin{equation}
D(q) = D^0(q) + D^0(q) \Pi^s(q) D(q).
\end{equation}
It can be summed up into geometric series, so one obtains

\begin{equation}
D(q) = {D^0(q) \over  1 - \Pi^s(q) D^0(q)}= D^0(q) + D^0(q)
\Pi^s(q) D^0(q) - D^0(q)\Pi^s(q)D^0(q) \Pi^s(q)D^0(q) + ... \ .
\end{equation}
Since $\Pi^s(q) = O(q^2)$ and $D^0(q) \sim (q^2)^{-1}$, the IR
singularity of the full photon propagator is determined by the IR
singularity of the free photon propagator, i.e., $D(q) = O
(D^0(q))$ with respect to the behavior at small photon momentum.

In fact, the current conservation condition (2.7), i.e., the
transversality of the photon self-energy lowers the quadratic
divergence of the corresponding integral (2.2) to a logarithmic
one. That is the reason why in QED only logarithmic divergences
survive. Thus in QED there is no mass gap and the relevant photon
SD equation is shown in Eq. (2.13). From it follows that the
behavior of the full gluon propagator at small gluon momentum is
determined by the behavior of its free PT counterpart. In other
words, in QED we can replace $\Pi(q)$ by its subtracted
counterpart $\Pi^s(q)$ from the very beginning ($\Pi(q)
\rightarrow \Pi^s(q)$), totally discarding the quadratically
divergent constant $\Delta^2(\lambda)$ from all the equations and
relations. The current conservation condition for the photon
self-energy (2.7), i.e., its transversality, and the condition
$q_{\mu}q_{\nu}D_{\mu\nu}(q) = i\xi$ (here and everywhere below
$\xi$ is the gauge-fixing parameter) imposed on the full gluon
propagator are consequences of gauge invariance. They should be
maintained at every stage of the calculations, since the photon is
a physical state. In other words, at all stages the current
conservation plays a crucial role in extracting physical
information from the $S$-matrix elements in QED, which are usually
proportional to the combination $j^{\mu}_1 (q)D_{\mu\nu}(q)
j^{\nu}_2(q)$. The current conservation condition $j^{\mu}_1 (q)
q_{\mu} = j^{\nu}_2(q)q_{\nu} =0$ implies that the unphysical
(longitudinal) component of the full photon propagator does not
change the physics of QED, i.e., only its physical (transversal)
component is important. In its turn this means that the
transversality condition imposed on the photon self-energy is also
important, because $\Pi_{\mu\nu}(q)$ itself is a correction to the
amplitude of the physical process, for example such as
electron-electron scattering.

Concluding, let us emphasize that the photon is always a massless
state, since in QED (unlike QCD, see below) the mass gap cannot be
realized.

\section{QCD}

For our purposes just like in QED it is convenient to begin with
the general description of the SD equation for the full gluon
propagator, and not for its inverse. Symbolically it can be
written down as follows:

\begin{equation}
D_{\mu\nu}(q) = D^0_{\mu\nu}(q) + D^0_{\mu\rho}(q) i
\Pi_{\rho\sigma}(q; D) D_{\sigma\nu}(q),
\end{equation}
where $D^0_{\mu\nu}(q)$ is the free gluon propagator.
$\Pi_{\rho\sigma}(q; D)$ is the gluon self-energy, and in general
it depends on the full gluon propagator due to non-abelian
character of QCD (see below as well). Thus the gluon SD equation
is highly NL, while the photon SD equation (2.1) is a linear one.
In what follows we omit the color group indices, since for the
gluon propagator (and hence for its self-energy) they are reduced
to the trivial $\delta$-function, for example $D^{ab}_{\mu\nu}(q)
= D_{\mu\nu}(q)\delta^{ab}$. Also, for convenience, we introduce
$i$ into the gluon SD equation (3.1).

In comparison with the photon self-energy (2.2), the gluon
self-energy $\Pi_{\rho\sigma}(q; D)$ is the sum of a few terms,

\begin{equation}
\Pi_{\rho\sigma}(q; D)=  \Pi^q_{\rho\sigma}(q) +
\Pi^{gh}_{\rho\sigma}(q) + \Pi_{\rho\sigma}^t(D) +
\Pi^{(1)}_{\rho\sigma}(q; D^2) + \Pi^{(2)}_{\rho\sigma}(q; D^4) +
\Pi^{(2')}_{\rho\sigma}(q; D^3),
\end{equation}
where $\Pi^q_{\rho\sigma}(q)$ describes the skeleton loop
contribution due to quark degrees of freedom (it is an analog of
the vacuum polarization tensor (2.2) in QED), while
$\Pi^{gh}_{\rho\sigma}(q)$ describes the skeleton loop
contribution due to ghost degrees of freedom. Both skeleton loop
integrals do not depend on the full gluon propagator $D$, so they
represent the linear contribution to the gluon SD equation.
$\Pi_{\rho\sigma}^t(D)$ represents the so-called constant skeleton
tadpole term. $\Pi^{(1)}_{\rho\sigma}(q; D^2)$ represents the
skeleton loop contribution, which contains the triple gluon
vertices only. $\Pi^{(2)}_{\rho\sigma}(q; D^4)$ and
$\Pi^{(2')}_{\rho\sigma}(q; D^3)$ describe topologically
independent skeleton two-loop contributions, which combine the
triple and quartic gluon vertices. The last four terms explicitly
contain the full gluon propagators in the corresponding powers
symbolically shown above, that is why they form the NL part of the
gluon SD equation. The analytical expressions for the
corresponding skeleton loop integrals \cite{10} (in which the
symmetry coefficients can be included) are of no importance here,
since we are not going to introduce into them any
truncations/approximations as well as to choose some special
gauge. Let us note that like in QED these skeleton loop integrals
are quadratically divergent at least in the PT, and therefore they
are assumed to be regularized (see remarks above and below).

\subsection{Subtractions}

Quite similar to the subtraction (2.3), let us formally subtract
from the full gluon self-energy (3.2) its value at zero point
(see, however, remarks below). Thus, one obtains

\begin{equation}
\Pi^s_{\rho\sigma}(q; D) = \Pi_{\rho\sigma}(q; D) -
\Pi_{\rho\sigma}(0; D) = \Pi_{\rho\sigma}(q; D) -
\delta_{\rho\sigma} \Delta^2 (\lambda; D).
\end{equation}
In this connection let us make in advance a few general remarks.
Contrary to QED, QCD being non-abelian gauge theory can suffer
from the IR singularities in the $q^2 \rightarrow 0$ limit due to
the self-interaction of massless gluon modes. Thus the initial
subtraction at zero point in the definition (3.3) may be dangerous
\cite{1}, indeed. That is why in all quantities below  the
dependence on the finite (slightly different from zero)
dimensionless subtraction point $\alpha$ is to be understood. In
other words, all the subtractions at zero and the Taylor
expansions around zero should be understood as the subtractions at
$\alpha$ and the Taylor expansions near $\alpha$, where they are
justified to use. From the technical point of view it is
convenient to put formally $\alpha=0$ in all derivations below,
and to restore the explicit dependence on non-zero $\alpha$ in all
quantities at the final stage only. However, we will restore the
explicit dependence on $\alpha$, when it will be necessary for
better understanding of the corresponding derivations (see section
IV below).

Let us remind once more that by mass gap we understand the
difference between the regularized gluon self-energy and its
subtracted (also regularized) counterpart. To demonstrate a
possible existence of a mass gap $\Delta^2 (\lambda; D)$ in QCD,
it is not important how $\lambda$ has been introduced and  how it
should be removed at the final stage. The mass gap itself is
mainly generated by the nonlinear interaction of massless gluon
modes, slightly corrected by the linear contributions coming from
the quark and ghost degrees of freedom, namely

\begin{equation}
\Delta^2 (\lambda; D)= \Pi^t(D) + \sum_a \Pi^a(0; D) =
\Delta^2_t(D) + \sum_a \Delta^2_a(0; D) ,
\end{equation}
where index "a" runs as follows: $a= q, gh, 1, 2, 2'$, and here,
obviously, the tensor indices are omitted. In these relation all
the quadratically divergent constants $\Pi^t(D)$ and $\Pi^a(0;
D)$, having the dimensions of a mass squared, are given by the
corresponding skeleton loop integrals at $q^2=0$, which appear in
Eq. (3.2). In this connection, let us remind that by the quadratic
divergences we conventionally understand the divergent constants
having the dimensions of a mass squared and summed up into the
mass gap (3.4). Then not losing generality, we can put
$\Delta^2(\lambda) = m^2 f(\lambda)$, where $m^2$ is some fixed
mass squared, and $f(\lambda)$ is some dimensionless function. Its
dependence on $\lambda$ is determined by the divergences of the
above-mentioned skeleton loop integrals. However, due to AF the
dependence is linear one (up to AF logarithm), so the divergence
becomes the quadratic one $\Delta^2(\lambda) \sim m^2 \lambda \sim
\Lambda^2$, indeed, like in the PT.

The subtracted gluon self-energy (3.3)

\begin{equation}
\Pi^s_{\rho\sigma}(q; D) \equiv \Pi^s(q; D) = \sum_a \Pi^s_a(q; D)
\end{equation}
is free from the tadpole contribution, because $\Pi^s_t(D) =
\Pi_t(D)- \Pi_t(D)=0$, by definition, at any $D$, while in the
gluon self-energy (3.2) it is explicitly present

\begin{equation}
\Pi_{\rho\sigma}(q; D) \equiv \Pi(q; D) = \Pi_t(D) + \sum_a
\Pi^s_a(q; D).
\end{equation}

The general decomposition of the subtracted gluon self-energy into
the independent tensor structures can be written down as follows:

\begin{equation}
\Pi^s_{\rho\sigma}(q; D) = T_{\rho\sigma}(q) q^2 \Pi^s(q^2; D) +
q_{\rho} q_{\sigma} \tilde{\Pi}^s(q^2; D),
\end{equation}
where both invariant functions $\Pi^s(q^2; D)$ and
$\tilde{\Pi}^s(q^2; D)$ are dimensionless functions of their
argument $q^2$. The subtracted gluon self-energy does not contain
the tadpole contribution, see Eq. (3.5). Let us note in advance
(see subsection B below) that in this case we can impose the color
current conservation condition on it, i.e., to put

\begin{equation}
q_{\rho} \Pi^s_{\rho\sigma}(q; D)= q_{\sigma}
\Pi^s_{\rho\sigma}(q; D) = 0,
\end{equation}
which implies $\tilde{\Pi}^s(q^2; D) = 0$. So the subtracted gluon
self-energy finally becomes the purely transversal

\begin{equation}
\Pi^s_{\rho\sigma}(q; D) = T_{\rho\sigma}(q) q^2 \Pi^s(q^2; D).
\end{equation}
Let us remind once more that we can expand $\Pi^s(q^2; D)$ in a
Taylor series near the subtraction point $\alpha$ at any $D$. Thus
the subtracted quantities are free from the quadratic divergences,
but the logarithmic ones at large $q^2$ can be still present in
$\Pi^s(q^2; D)$, like in QED.

Concluding, let us note that we are not going to impose the
transversality condition on the gluon self-energy (3.6) itself
(see below). That is why we need no its decomposition into the
independent tensor structures.

\subsection{A self-consistent violation of color gauge
invariance/symmetry (SCVCGI/S)}

QCD is $SU(3)$ color gauge invariant theory, but however:

{\bf (i)}. Due to color confinement, the gluon (unlike the photon)
is not a physical state. Moreover, there is no such physical
amplitude to which the gluon self-energy (like the photon
self-energy) may directly contribute. For example, quark/quark and
quark/antiquark scattering are not a physical processes.

{\bf (ii)}. Contrary to the conserved currents in QED, the color
conserved currents do not play any role in the extraction of
physical information from the $S$-matrix elements for the
corresponding physical processes and quantities in QCD. In other
words, not the conserved color currents, but only their
color-singlet counterparts, which can even be partially conserved,
contribute directly to the $S$-matrix elements describing this or
that physical process/quantity. For example, such an important
physical QCD parameter as the pion decay constant is given by the
following $S$-matrix element: $<0|J^i_{5\mu}(0)|\pi^j(q)>= i
q_{\mu} F_{\pi} \delta^{ij}$, where $J^i_{5\mu}(0)$ is just the
axial-vector current, while $|\pi^j(q)>$ describes the pion
bound-state amplitude, and $i, j$ are flavor indices.

{\bf (iii)}. Moreover, in QCD (contrary to QED) exists a direct
evidence/indication that the transversality of the full gluon
self-energy may be violated beyond the PT, indeed.

The color gauge invariance condition for the full gluon
self-energy (3.2) can be reduced to the three independent
transversality conditions imposed on it. It is well known that the
quark contribution can be made transversal independently of the
pure gluon contributions within any regularization scheme which
preserves gauge invariance, for example such as the dimensional
regularization method (DRM) \cite{11} (see Refs. \cite{1,2,8,9} as
well). So, we can put

\begin{equation}
q_{\rho} \Pi^q_{\rho\sigma}(q) =q_{\sigma} \Pi^q_{\rho\sigma}(q) =
0.
\end{equation}
Explicitly it can be shown in lower order of the PT (see, for
example Refs. \cite{2,6,7,8,9}). It is assumed, however, that in
principle it should be valid in every order of the PT, thus going
beyond the PT.

In the same way the sum of the gluon contributions can be made
transversal by taking into account the ghost contribution, so
again one can put

\begin{equation}
q_{\rho} \Bigl[ \Pi^{(1)}_{\rho\sigma}(q; D^2) +
\Pi^{(2)}_{\rho\sigma}(q; D^4) + \Pi^{(2')}_{\rho\sigma}(q; D^3) +
\Pi^{gh}_{\rho\sigma}(q) \Bigr] = 0.
\end{equation}
The role of ghost degrees of freedom is to cancel the unphysical
(longitudinal) component of gauge boson (gluon) propagator in
every order of the PT, i.e., going beyond the PT and thus being
the general one. The previous general condition of cancellation
(3.11) just demonstrates this, since it contains the corresponding
skeleton loop integrals. As in a quark case, the explicit
cancellation can be shown, nevertheless, only in lower order of
the PT. For this we should put $D = D^0 \equiv D_0$ and
approximate all other quantities, entering the corresponding
skeleton loop integrals in the relation (3.11), by their free PT
counterparts (see, for example Refs. \cite{2,8,9}).

However, there is no such regularization scheme (preserving or not
gauge invariance) in which the transversality condition for the
full gluon self-energy could be satisfied unless the constant
skeleton tadpole term

\begin{equation}
\Pi_t(D) =  g^2 \int {i d^4 q_1 \over (2 \pi)^4} T^0_4 D(q_1),
\end{equation}
is disregarded from the very beginning (here $T^0_4$ is the
four-gluon point-like vertex and we omit the tensor and color
indices, as unimportant for further purpose). It is nothing else
but the quadratically divergent in the PT constant. It explicitly
violets the transversality condition for the full gluon
self-energy, since formally $q_{\rho} \Pi^t_{\rho\sigma}(D) =
q_{\rho} \delta_{\rho\sigma} \Delta^2_t(D) = q_{\sigma}
\Delta^2_t(D) \neq 0$. In the PT, when the full gluon propagator
is always approximated by the free one, the constant tadpole term
is set to be zero within the DRM \cite{2,8,9}, i.e.,
$\Pi^t_{\rho\sigma}(D_0) =0$. So in the PT the transversality
condition for the full gluon self-energy is always satisfied.
However, even in the DRM this is not an exact result, but rather
an embarrassing prescription, as pointed out in Ref. \cite{8}. To
show explicitly that even in the PT there are still problems, it
is instructive to substitute the first iteration of the gluon SD
equation (3.1) into the previous expression. Symbolically it looks
like $D = D_0 + D_0i\Pi(D_0)D_0 + ...$, where we omit all the
indices. Doing so, one obtains

\begin{eqnarray}
\Pi_t(D) &=& \Pi_t(D_0) + g^2 \int {i d^4 q_1 \over (2 \pi)^4}
T^0_4 [D_0(q_1)]^2 i \Pi(q_1; D_0) +...
\nonumber\\
&=& \Pi_t(D_0) + \Pi(0; D_0) g^2 \int {i^2 d^4 q_1 \over (2
\pi)^4} T^0_4 [D_0(q_1)]^2 + g^2 \int {i^2 d^4 q_1 \over (2
\pi)^4} T^0_4 [D_0(q_1)]^2 q^2_1
\Pi^s(q^2_1; D_0) +... \ . \nonumber\\
\end{eqnarray}
Here we introduce the subtraction as follows: $\Pi^s(q_1; D_0) =
\Pi(q_1; D_0) - \Pi(0; D_0)$, and  $\Pi(0; D_0)= \Pi_t(D_0) +
...$, where all other quadratically divergent constant terms are
omitted, for simplicity. In the second line of Eq. (3.13) the
first integral is not only UV divergent but it is IR singular as
well. If now we omit the first term in accordance with the
above-mentioned prescription, the product of this integral and the
tadpole term $\Pi_t(D_0)$ remains, nevertheless, undetermined.
Moreover, the structure of the second integral in this line is
completely different from the divergent constant integral
$\Pi_t(D_0)$. This constant term is also not determined, since in
general we do not know the behavior of $\Pi^s(q^2_1; D_0)$  at
small and large $q_1^2$. All this reflects the general problem
that all such kind of massless integrals $\int ( d^d q / (2
\pi)^d) (q_{\mu_1}...q_{\mu_p} / (q^2)^n)$ are ill defined, since
there is no dimension where they are meaningful; they are either
IR singular or UV divergent \cite{8}. This prescription clearly
shows that the DRM, preserving gauge invariance, nevertheless,
does not alone provide us insights into the correct treatment of
the power-like IR singularities (we will address this problem in
the second (II) part of our investigation). However, in the PT we
can adhere to the prescription that such massless tadpole
integrals can be discarded in the DRM \cite{8}. This is the only
way for ghosts to validate the transversality condition for the
full gluon self-energy in PT QCD. It makes the full gluon
propagator the purely transversal. Then the $S$-matrix elements
for physical quantities and processes in PT QCD become free from
unphysical degrees of freedom of gauge bosons, maintaining thus
the unitarity of $S$-matrix in this theory.

So, we conclude that beyond the PT, i.e., in the general case the
transversality of the full gluon self-energy may be violated. In
other words, in the general case, i.e., beyond the PT, we cannot
discard the tadpole term (3.12) from the very beginning. If we do
not know how to treat it, this does not mean that we should
neglect it at all. Its regularized version should be explicitly
taken into account. Moreover, all other quadratically divergent
constants, which have been summed up into the mass gap (3.4),
cannot be discarded like in QED, since the transversality
condition for the gluon self-energy is not going to be imposed
(see below). In other words, in QCD the quadratic divergences of
the corresponding skeleton loop integrals cannot be lowered to the
logarithmic ones, and therefore the mass gap (3.4) should be
explicitly taken into account in this theory.

Thus in order to realize the mass gap (3.4) our proposal is not to
impose the transversality condition on the gluon self-energy
(3.2), i.e., to admit that in general case

\begin{equation}
q_{\rho} \Pi_{\rho\sigma}(q; D) =q_{\sigma} \Pi_{\rho\sigma}(q; D)
\neq 0,
\end{equation}
indeed. At the same time, we would like to preserve the color
gauge invariance condition for the full gluon propagator, i.e.,
within our approach the relation

\begin{equation}
q_{\mu}q_{\nu}D_{\mu\nu}(q) = i \xi,
\end{equation}
which is nothing else but the ST identity, always holds. This is
important for the renormalization.

The lesson which comes from QED is that if one preserves the
transversality of the photon self-energy at every stage, then
there is no mass gap. Thus in order to realize a mass gap in QCD,
our proposal is not to impose the transversality condition on the
gluon self-energy, Eq. (3.14), but preserving the ST identity for
the full gluon propagator, Eq. (3.15).

Concluding, a few general remarks are in order:

{\bf 1}. We would like to emphasize the special role of the
constant skeleton tadpole term (3.12) in the NP QCD dynamics. Its
existence is a direct evidence that the transversality of the
gluon self-energy may be violated beyond the PT theory.

{\bf 2}. The second important observation is that the ghosts
themselves cannot now automatically provide the transversality of
the gluon propagator in NP QCD. Thus, we sacrifice the general
role of ghosts in order to realize a mass gap. To realize a mass
gap is much more necessary than to maintain the general role of
ghosts. At long last, the role of ghosts is mainly kinematical,
while the mass gap dominates the dynamics of QCD at large distance
(see below). This is important for understanding of the
confinement mechanism.

{\bf 3}. However, let us note in advance that how to restore the
transversality of the full gluon propagator relevant for NP QCD
will be explained in part II of our investigation. In other words,
at the initial stage we violate the transversality of the gluon
self-energy in order to realize a mass gap, while restoring the
transversality of the full gluon propagator relevant for NP QCD at
the final stage. At the same time, the ST identity for the gluon
propagator is always valid within our approach. All this will make
it possible to maintain the unitarity of the $S$-matrix in NP QCD.

{\bf 4}. The discussion above does not mean that we need no ghosts
at all. We need them in other sectors of QCD, for example in the
quark-gluon ST identity, which contains the so-called ghost-quark
scattering kernel explicitly \cite{1}. It provides an important
piece of information on quark degrees of freedom themselves. If
one omits the ghosts, then it will be totally lost (for details
see Ref. \cite{12,13,14}, a recent publication \cite{15} and
references therein).

{\bf 5}. The transversality condition for the gluon self-energy
can be satisfied partially, i.e., if one imposes it on quark and
gluon (along with ghost) degrees of freedom, as it follows from
the relations (3.10) and (3.11). Then the mass gap is to be
reduced to $\Pi^t(D)$, since all other constants $\Pi^a(0;D)$ can
be discarded in this case, see Eq. (3.4). However, we will stick
to our proposal not to impose the transversality condition on the
gluon self-energy at all, and thus to deal with the mass gap on
account of all the possible contributions.

\subsection{General structure of the gluon SD equation}

Our strategy is not to impose the transversality condition on the
gluon self-energy in order to realize a mass gap despite whether
or not the tadpole term is explicitly present. At the same time,
we would like to preserve the ST identity (3.15), as underlined
above. It implies that the general tensor decomposition of the
full gluon propagator becomes the standard one, namely

\begin{equation}
D_{\mu\nu}(q) = i \left\{ T_{\mu\nu}(q) d(q^2) + \xi L_{\mu\nu}(q)
\right\} {1 \over q^2},
\end{equation}
where $d(q^2) \equiv d(q^2; \xi)$ is the full gluon invariant
function (the full gluon form factor or equivalently the full
effective charge ("running")). To show that our strategy works,
let us substitute the subtraction (3.3), on account of the
relation (3.9), into the initial gluon SD equation (3.1). Then one
obtains

\begin{equation}
D_{\mu\nu}(q) = D^0_{\mu\nu}(q) + D^0_{\mu\rho}(q)i
T_{\rho\sigma}(q) q^2 \Pi^s(q^2; D)D_{\sigma\nu}(q) +
D^0_{\mu\sigma}(q)i \Delta^2(\lambda; D) D_{\sigma\nu}(q).
\end{equation}
In the presence of a mass gap, it is instructive to introduce the
general tensor decomposition of the auxiliary free gluon
propagator as follows:

\begin{equation}
D^0_{\mu\nu}(q)=i[ T_{\mu\nu}(q) + L_{\mu\nu}(q) d_0(q^2)](1/q^2).
\end{equation}
Evidently, any temporary deviation in the auxiliary free gluon
propagator from the standard free gluon propagator in the presence
of a mass gap may appear only in its unphysical (longitudinal)
component. There is no dynamics in any free gluon propagator
provided by the nontrivial (not equal to one) form factor,
affiliated with its transversal component.

Substituting all these decompositions (3.16) and (3.18) into the
gluon SD equation (3.17), one obtains

\begin{equation}
d(q^2) = {1 \over 1 + \Pi^s(q^2; D) + (\Delta^2(\lambda; D) /
q^2)},
\end{equation}
and

\begin{equation}
d_0(q^2) = {\xi \over 1 - \xi (\Delta^2(\lambda; D) / q^2)}.
\end{equation}
However, we need the standard free gluon propagator rather than
its auxiliary counterpart, despite the latter one being reduced to
the former one in the formal PT $\Delta^2 (\lambda; D)=0$ limit.
To achieve this goal, the auxiliary free gluon propagator defined
in Eqs. (3.18) and (3.20) is to be equivalently replaced as
follows:

\begin{equation}
D^0_{\mu\nu}(q) \Longrightarrow D^0_{\mu\nu}(q) + i \xi
L_{\mu\nu}(q) d_0(q^2) { \Delta^2(\lambda; D) \over q^4},
\end{equation}
where $D^0_{\mu\nu}(q)$ in the right-hand-side is the standard
free gluon propagator now, i.e.,

\begin{equation}
D^0_{\mu\nu}(q) = i \left\{ T_{\mu\nu}(q) + \xi L_{\mu\nu}(q)
\right\} {1 \over q^2}.
\end{equation}
Then the gluon SD equation in the presence of a mass gap (3.17),
after the replacement (3.21) and doing some algebra, on account of
the explicit expression for the auxiliary free gluon form factor
(3.20), becomes

\begin{eqnarray}
D_{\mu\nu}(q) &=& D^0_{\mu\nu}(q) + D^0_{\mu\rho}(q)i
T_{\rho\sigma}(q) q^2 \Pi^s(q^2; D) D_{\sigma\nu}(q) \nonumber\\
&+& D^0_{\mu\sigma}(q)i \Delta^2(\lambda; D) D_{\sigma\nu}(q) + i
\xi^2 L_{\mu\nu}(q) { \Delta^2(\lambda; D) \over q^4}.
\end{eqnarray}
Here and from now on $D^0_{\mu\nu}(q)$ is the standard free gluon
propagator (3.22). Using it explicitly, and on account of the
decomposition (3.16), this equation can be further simplified to

\begin{equation}
D_{\mu\nu}(q) = D^0_{\mu\nu}(q) - T_{\mu\sigma}(q)
\Bigl[\Pi^s(q^2; D) + { \Delta^2(\lambda; D) \over q^2} \Bigr]
D_{\sigma\nu}(q).
\end{equation}

It is easy to check that the full gluon propagator given by this
equation (and hence by Eq. (3.23)) satisfies the ST identity
(3.15), indeed, even in the presence of a mass gap. So the full
gluon propagator is the expression (3.16) with the full gluon form
factor given in Eq. (3.19), which obviously satisfies Eqs. (3.17),
(3.23) and (3.24) simultaneously. The only price we have paid so
far is the gluon self-energy, while its subtracted counterpart is
always transversal. At the same time, the full gluon propagator
(3.16) and the free PT gluon propagator (3.22) automatically
satisfy the ST identity (3.15). So, one can conclude that our
mechanism for the realization of the mass gap is rather
self-consistent.

Let us emphasize that the expression for the full gluon form
factor shown in the relation (3.19) cannot be considered as the
formal solution for the full gluon propagator $D$ (see Eq.
(3.16)), since both the mass gap $\Delta^2(\lambda; D)$ and the
invariant function $\Pi^s(q^2; D)$ depend on $D$ themselves. It
clearly follows from the relation (3.19) that the effect of the
mass gap dominates the IR region when the gluon momentum goes to
zero, and this effect vanishes when the gluon momentum goes to
infinity. This once more underlines a close intrinsic link between
the NP dynamics governed by the mass gap and the structure of the
true QCD vacuum at large distances ($q^2 \rightarrow 0$). It is
worth recalling once more that in the opposite limit, i.e., at
large $q^2$, the subtracted gluon self-energy $\Pi^s(q^2; D)$ may
still suffer from the logarithmic divergences, like in QED.

In the formal PT $\Delta^2(\lambda; D)=0$ limit, from the gluon SD
equation (3.23) one recovers the standard gluon SD equation (3.1),
and the gluon self-energy coincides with its subtracted
counterpart like in QED, see Eq. (3.3). Then the formal "solution"
(3.19) will not depend on the mass gap. The above-mentioned
general role of ghosts is to be automatically restored. They will
again provide the cancellation of the longitudinal component of
the full gluon propagator in this limit.

Concluding, we have established the general structure of the full
gluon propagator in the presence of a mass gap. Moreover, we have
explicitly shown that the initial gluon SD equation (3.17) has the
same "solution" as the final gluon SD equation (3.23), that is Eq.
(3.16) along with the relation (3.19).

\section{Nonlinear iteration solution}

In order to find a formal solution for the regularized full gluon
propagator (3.16), on account of its effective charge (3.19) (or
equivalently the full gluon form factor), let us rewrite the
latter one in the form of the corresponding transcendental (i.e.,
not algebraic) equation, namely

\begin{equation}
d(q^2) = 1 - \Bigl[ \Pi(q^2; d) + {\Delta^2(\lambda; d) \over q^2}
\Bigr] d(q^2) = 1 - P(q^2; d) d(q^2),
\end{equation}
suitable for the formal nonlinear iteration procedure. Here we
replace the dependence on $D$ by the equivalent dependence on $d$.
Also, for simplicity, we replaced $\Pi^s(q^2; d) \rightarrow
\Pi(q^2; d)$. For future purposes, it is convenient to introduce
short-hand notations as follows:

\begin{eqnarray}
\Delta^2(\lambda; d=d^{(0)} + d^{(1)} + d^{(2)} + ... + d^{(m)}+
...
) &=& \Delta^2_m = \Delta^2 c_m(\lambda, \alpha, \xi, g^2)
\equiv c_m \Delta^2, \nonumber\\
\Pi(q^2; d=d^{(0)} + d^{(1)}+d^{(2)}+ ... + d^{(m)} + ...) &=&
\Pi_m(q^2),
\end{eqnarray}
and

\begin{equation}
P_m(q^2) = \Bigl[ \Pi_m(q^2) + {\Delta^2_m \over q^2} \Bigr], \
m=0,1,2,3,... \ .
\end{equation}
In these relations $\Delta^2_m$ are the auxiliary mass squared
parameters, while $\Delta^2 \equiv \Delta^2(\lambda; d)$ is the
mass gap itself. Via the corresponding subscripts the
dimensionless constants $c_m$ depend on which iteration for the
gluon form factor $d$ is actually done. They may depend on the
dimensionless coupling constant squared $g^2$, as well as on the
gauge-fixing parameter $\xi$. We also introduce in advance the
explicit dependence on the finite (slightly different from zero)
dimensionless subtraction point $\alpha$, as pointed out above.
The dependence of $\Delta^2$ on all these parameters, as well as
on the number of different flavors $N_f$ and colors $N_c$, is not
shown explicitly, and if necessary it can be restored any time.
Let us also recall that all the invariant functions $\Pi_m(q^2)$
can be expand in a formal Taylor series near the finite
subtraction point $\alpha$. If it were possible to express the
full gluon form factor $d(q^2)$ in terms of these quantities then
it would be the formal solution for the full gluon propagator. In
fact, this is nothing but the skeleton loops expansion, since the
regularized skeleton loop integrals, contributing to the gluon
self-energy, have to be iterated. This is the so-called general
nonlinear iteration solution. As mentioned above, no
truncations/approximations/assumptions and no special gauge choice
have been made. This formal expansion is not a PT series. The
magnitude of the coupling constant squared and the dependence of
the regularized skeleton loop integrals on it is completely
arbitrary. Let us emphasize once more that through this paper the
mass gap $\Delta^2 \equiv \Delta^2(\lambda, \alpha, \xi, g^2)$ is
a "bare" one, i.e., it is only regularized (with the help of
$\lambda$ and $\alpha$) in order to assign a mathematical meaning
to all derivations involving it.

It is instructive to describe the general iteration procedure in
some details. Evidently, $d^{(0)}=1$, and this corresponds to the
approximation of the full gluon propagator by its free
counterpart. Doing the first iteration in Eq. (4.1), one thus
obtains

\begin{equation}
d(q^2) = 1 - P_0(q^2) + ... = 1 + d^{(1)}(q^2) + ...,
\end{equation}
where obviously

\begin{equation}
d^{(1)}(q^2) = - P_0(q^2).
\end{equation}
Carrying out the second iteration, one gets

\begin{equation}
d(q^2) = 1 - P_1(q^2) [ 1 + d^{(1)}(q^2) ] + ... = 1 +
d^{(1)}(q^2) + d^{(2)}(q^2) + ...,
\end{equation}
where

\begin{equation}
d^{(2)}(q^2) = - d^{(1)}(q^2) - P_1(q^2) [ 1 - P_0(q^2)].
\end{equation}
Doing the third iteration, one further obtains

\begin{equation}
d(q^2) = 1 - P_2(q^2) [ 1 + d^{(1)}(q^2) + d^{(2)}(q^2)] + ... = 1
+ d^{(1)}(q^2) + d^{(2)}(q^2) + d^{(3)}(q^2) + ...,
\end{equation}
where

\begin{equation}
d^{(3)}(q^2) = - d^{(1)}(q^2) - d^{(2)}(q^2) - P_2(q^2) [ 1 -
P_1(q^2)(1 - P_0(q^2))],
\end{equation}
and so on for the next iterations.

Thus up to the third iteration, one finally arrives at

\begin{equation}
d(q^2) = \sum_{m=0}^{\infty} d^{(m)}(q^2) = 1 - \Bigl[ \Pi_2(q^2)
+ {\Delta^2_2 \over q^2} \Bigr] \Bigl[ 1 - \Bigl[ \Pi_1(q^2) +
{\Delta^2_1 \over q^2} \Bigr] \Bigl[ 1 - \Pi_0(q^2) - {\Delta^2_0
\over q^2} \Bigr] \Bigr] + ... \ .
\end{equation}
We restrict ourselves by the iterated gluon form factor up to the
third term, since this already allows to show explicitly some
general features of the nonlinear iteration solution.

\subsection{Splitting/shifting procedure}

Doing some tedious algebra, the previous expression (4.10) can be
rewritten as follows:

\begin{eqnarray}
d(q^2) &=& \Bigl[ 1 - \Pi_2(q^2) + \Pi_1(q^2) \Pi_2(q^2) -
\Pi_0(q^2)
\Pi_1(q^2)\Pi_2(q^2) + ... \Bigr] \nonumber\\
&+& {1 \over q^2} \Bigl[ \Pi_2(q^2)\Delta^2_1 +
\Pi_1(q^2)\Delta^2_2 - \Pi_0(q^2) \Pi_1(q^2)\Delta^2_2 -
\Pi_0(q^2) \Pi_2(q^2)\Delta^2_1
- \Pi_1(q^2) \Pi_2(q^2)\Delta^2_2 + ... \Bigr] \nonumber\\
&-& {1 \over q^4} \Bigl[ \Pi_0(q^2) \Delta^2_1 \Delta^2_2 +
\Pi_1(q^2) \Delta^2_0 \Delta^2_2 + \Pi_2(q^2) \Delta^2_0
\Delta^2_1 + ... \Bigr] \nonumber\\
&-& {1 \over q^2} \Bigl[ \Delta^2_2 -  {\Delta^2_1 \Delta^2_2
\over q^2} + { \Delta^2_0 \Delta^2_1 \Delta^2_2 \over q^4} + ...
\Bigr].
\end{eqnarray}
This formal expansion contains three different types of terms. The
first type are the terms which contain only different combinations
of $\Pi_m(q^2)$ (they are not multiplied by inverse powers of
$q^2$); the third type of terms contains only different
combinations of $(\Delta^2_m / q^2)$. The second type of terms
contains the so-called mixed terms, containing the first and third
types of terms in different combinations. The two last types of
terms are multiplied by the corresponding powers of $1/q^2$. Such
structure of terms will be present in each iteration term for the
full gluon form factor. However, any of the mixed terms can be
split exactly into the first and third types of terms. For this
purpose the formal Taylor expansions for $\Pi_m(q^2)$ around the
finite subtraction point $\alpha$ should be used. Thus an exact IR
structure of the full gluon form factor (which just is our primary
goal to establish) is determined not only by the third type of
terms. It gains contributions from the mixed terms as well, but
without changing its functional dependence (see remarks below). To
demonstrate this in some detail, it is convenient to express the
previous expansion (4.11) in terms of dimensionless variable and
parameters, namely

\begin{equation}
x = {q^2 \over M^2}, \quad c = {\Delta^2 \over M^2}, \quad \alpha
= {\mu^2 \over M^2},
\end{equation}
where $M^2$ is some auxiliary fixed mass squared, and $\mu^2$ is
the point close to $q^2=0$ (to be not mixed up with the tensor
index). Also, in the formal PT $\Delta^2 =0$ limit $c \equiv
c(\lambda, \alpha, \xi, g^2)=0$ (unlike $c_m$ in the relations
(4.2)), since $M^2$ is fixed. Using further relations (4.2), and
on account of the relations (4.12), the expansion (4.11) becomes

\begin{eqnarray}
d(x) &=& \Bigl[ 1 - \Pi_2(x) + \Pi_1(x) \Pi_2(x) - \Pi_0(x)
\Pi_1(x)\Pi_2(x) + ... \Bigr] \nonumber\\
&+& {c \over x} \Bigl[ \Pi_2(x) c_1  + \Pi_1(x)c_2  - \Pi_0(x)
\Pi_1(x) c_2  - \Pi_0(x) \Pi_2(x) c_1
- \Pi_1(x) \Pi_2(x) c_2  + ... \Bigr] \nonumber\\
&-& {c^2 \over x^2} \Bigl[ \Pi_0(x) c_1 c_2  + \Pi_1(x) c_0 c_2
+ \Pi_2(x) c_0 c_1  + ... \Bigr] \nonumber\\
&-& {c \over x} \Bigl[ c_2  -  {c_1 c_2 c \over x} + { c_0 c_1 c_2
c^2 \over x^2} + ... \Bigr].
\end{eqnarray}

Taking into account the above-mentioned formal Taylor expansions

\begin{equation}
\Pi_m(x) = \sum_{n=0}^{\infty} (x - \alpha)^n \Pi^{(n)}_m (\alpha)
= \sum_{n=0}^{\infty} \Bigl[ \sum_{k=0}^n p_{nk} x^k \alpha^{n-k}
\Bigr] \Pi^{(n)}_m (\alpha),
\end{equation}
for example, the mixed term $(c_1 c / x) \Pi_2(x)$ can be then
exactly split/decomposed as follows:

\begin{equation}
{c_1 c \over x} \Pi_2(x) = {c_1 c \over x} \sum_{n=0}^{\infty}
\Bigl[ \sum_{k=0}^n p_{nk} x^k \alpha^{n-k} \Bigr] \Pi^{(n)}_2
(\alpha) = \Bigl( {c \over x} \Bigr) P_1(\alpha) + P_0(\alpha) +
O_2(x).
\end{equation}
Here and below the dependence on all other possible parameters is
not shown, for simplicity. The dimensionless function $O_2(x)$ is
of the order $x$ at small $x$; otherwise it remains arbitrary. The
first term now is to be shifted to the third type of terms, while
the remaining terms are to be shifted to the first type of terms.
All other mixed terms of similar structure should be treated
absolutely in the same way.

The mixed term $(c_1 c_2 c^2 / x^2) \Pi_0(x)$ can be split as

\begin{equation}
{c_1 c_2 c^2 \over x^2} \Pi_0(x) = {c_1 c_2 c^2 \over x^2}
\sum_{n=0}^{\infty} \Bigl[ \sum_{k=0}^n p_{nk} x^k \alpha^{n-k}
\Bigr] \Pi^{(n)}_0 (\alpha) = \Bigl( {c \over x} \Bigr)^2
P_2(\alpha) + \Bigl( {c \over x} \Bigr) N_1(\alpha) + N_0(\alpha)
+ O_0(x),
\end{equation}
where the dimensionless function $O_0(x)$ is of the order $x$ at
small $x$; otherwise it remains arbitrary. Again the first two
terms should be shifted to the third type of terms, while the last
two terms should be shifted to the first type of terms.

Similarly to the formal Taylor expansion (4.14), we can write

\begin{equation}
\Pi_m(x)\Pi_{m'} (x)= \Pi_{mm'} (x) = \sum_{n=0}^{\infty} (x -
\alpha)^n \Pi^{(n)}_{mm'} (\alpha) = \sum_{n=0}^{\infty} \Bigl[
\sum_{k=0}^n p_{nk} x^k \alpha^{n-k} \Bigr] \Pi^{(n)}_{mm'}
(\alpha).
\end{equation}
Then, for example the mixed term $(c_2 c /x) \Pi_0(x) \Pi_1(qx)$
can be split as

\begin{equation}
{ c_2 c \over x} \Pi_0(x) \Pi_1(x) = { c_2 c \over x} \Pi_{01}(x)=
{ c_2 c \over x} \sum_{n=0}^{\infty} \Bigl[ \sum_{k=0}^n p_{nk}
x^k \alpha^{n-k} \Bigr] \Pi^{(n)}_{01} (\alpha)= \Bigl( { c \over
x} \Bigr) M_1(\alpha) + M_0(\alpha) + O_{01}(x),
\end{equation}
where the dimensionless function $O_{01}(x)$ is of the order $x$
at small $x$; otherwise it remains arbitrary. Again the first term
should be shifted to the third type of terms, while other two
terms are to be shifted to the first type of terms.

Completing this exact splitting/shifting procedure in the
expansion (4.13), and restoring the explicit dependence on the
dimensional variable and parameters (4.12), one can equivalently
present the initial expansion (4.11) as follows:

\begin{equation}
d(q^2) = \Bigl( {\Delta^2 \over q^2} \Bigr) B_1(\lambda, \alpha,
\xi, g^2) + \Bigl( {\Delta^2 \over q^2} \Bigr)^2 B_2(\lambda,
\alpha, \xi, g^2) + \Bigl( {\Delta^2 \over q^2} \Bigr)^3
B_3(\lambda, \alpha, \xi, g^2) + ... + d_3(q^2; \Delta^2) + ... \
,
\end{equation}
where we use notations (4.2) explicitly now, since the
coefficients of the above-used expansions depend in general on the
same set of parameters: $\lambda, \alpha, \xi, g^2$. The invariant
function $d_3(q^2; \Delta^2)$ is dimensionless and it is free from
the power-type IR singularities; otherwise it remains arbitrary.
We have restored the dependence on the mass gap $\Delta^2$ instead
the dependence on the parameter $c$. In the formal PT $\Delta^2=0$
limit it survives, and is to be reduced to the sum of the first
type of terms in the expansion (4.11). The generalization on the
next iterations is almost obvious.

Concluding, let us underline that the splitting/shifting procedure
does not change the structure of the nonlinear iteration solution
at small $q^2$. It only changes the coefficients at inverse powers
of $q^2$ in the corresponding expansion. In other words, it makes
it possible to rearrange the terms in the initial expansion (4.11)
in order to get it in the final form (4.19). Also, in the $q^2
\rightarrow 0$ limit, it is legitimated to suppress the subtracted
gluon self-energy in comparison with the mass gap term in the
initial Eq. (4.1). Nevertheless, as a result of the
splitting/shifting procedure, which becomes almost trivial in this
case, one will obtain the same expansion (4.19) with only
different residues, as just mentioned above, and with $d_3(q^2;
\Delta^2)=1$ in this case.

\subsection{The exact structure of the general nonlinear iteration solution}

Substituting the generalization of the expansion (4.19) on all
iterations and doing some algebra, the general nonlinear iteration
solution for the regularized full gluon propagator (3.16) can be
exactly decomposed as the sum of the two principally different
terms as follows:

\begin{equation}
D_{\mu\nu}(q) = i \left\{ T_{\mu\nu}(q) d(q^2; \xi) + \xi
L_{\mu\nu}(q) \right\} {1 \over q^2} = D^{INP}_{\mu\nu}(q;
\Delta^2)+ D^{PT}_{\mu\nu}(q; \Delta^2),
\end{equation}
so that $D_{\mu\nu}(q) \equiv D_{\mu\nu}(q; \Delta^2)$. Here

\begin{equation}
D^{INP}_{\mu\nu}(q; \Delta^2) = i T_{\mu\nu}(q) d^{INP}(q^2;
\Delta^2) {1 \over q^2} = i T_{\mu\nu}(q) {\Delta^2 \over (q^2)^2}
L (q^2; \Delta^2),
\end{equation}

\begin{equation}
L(q^2; \Delta^2) = \sum_{k=0}^{\infty} \Bigl( {\Delta^2 \over q^2}
\Bigr)^k \Phi_k(\lambda, \alpha, \xi, g^2)= \sum_{k=0}^{\infty}
\Bigl( {\Delta^2 \over q^2} \Bigr)^k \sum_{m=0}^{\infty}
\Phi_{km}(\lambda, \alpha, \xi, g^2),
\end{equation}
while

\begin{equation}
D^{PT}_{\mu\nu}(q; \Delta^2) = i \Bigr[ T_{\mu\nu}(q) d^{PT}(q^2;
\Delta^2) + \xi L_{\mu\nu}(q) \Bigl] {1 \over q^2}.
\end{equation}

In Eqs. (4.20) and (4.21) the superscript "INP" stands for the
intrinsically NP part of the full gluon propagator. Let us
emphasize that the general problem of convergence of formal (but
regularized) series in Eq.(4.20) is irrelevant here. In other
words, it does not make any sense to discuss the convergence of
such kind of series before the renormalization program is
performed (which will allow to see whether the mass gap survives
it at all or not). Anyway, the problem how to remove the UV
overlapping divergences (Ref. \cite{16} and for a more detail
remarks see appendix C) and usual overall ones \cite{1,2,7,8,9} is
a standard one. Our problem will be how to deal with severe IR
singularities due to their novelty and genuine NP character.
Fortunately, there already exists a well-elaborated mathematical
formalism for this purpose, namely the distribution theory (DT)
\cite{17} into which the DRM \cite{11} should be correctly
implemented (see also Refs. \cite{18,19}). It will be explicitly
used in part II of our investigation.

We distinguish between the two terms in Eq. (4.20) {\bf first} of
all by the character of the corresponding IR singularities, when
the gluon momentum goes to zero. {\bf Secondly}, by the explicit
presence of the mass gap, when it formally goes to zero then the
INP term vanishes, while the PT term survives. The {\bf third}
point is that the INP part depends only on the transversal degrees
of freedom of gauge bosons, while the PT part has the longitudinal
component as well.

The INP part of the full gluon propagator is characterized by the
presence of severe power-type (or equivalently NP) IR
singularities $(q^2)^{-2-k}, \ k=0,1,2,3,...$. So these IR
singularities are defined as more singular than the power-type IR
singularity of the free gluon propagator $(q^2)^{-1}$, which thus
can be defined as the PT IR singularity. The INP part of the full
gluon propagator (4.21), apart from the structure $(\Delta^2 /
q^4)$, is nothing but the corresponding Laurent expansion
(explicitly shown in Eq. (4.22)) in integer powers of $q^2$
accompanied by the corresponding powers of the mass gap squared
and multiplied by the $q^2$-independent factors, the so-called
residues $\Phi_k(\lambda, \alpha, \xi, g^2) = \sum_{m=0}^{\infty}
\Phi_{km}(\lambda, \alpha, \xi, g^2)$. The sum over $m$ indicates
that an infinite number of iterations (all iterations) of the
corresponding regularized skeleton loop integrals invokes each
severe IR singularity labelled by $k$. It is worth emphasizing
that now this Laurent expansion cannot be summed up into anything
similar to the initial Eq. (3.19), since its residues at poles
gain additional contributions due to the splitting/shifting
procedure, i.e., they become arbitrary. However, this
arbitrariness is not a problem. The functional dependence, which
has been established exactly, is all that matters (a correct
treatment of severe IR singularities will be given in part II of
our investigation, as mentioned above). It is worth emphasizing
once more that just the mass gap term in Eq. (4.1) determines the
functional structure of the INP part (4.21)-(4.22) of the full
gluon propagator.

Thus the true QCD vacuum is really beset with severe IR
singularities. Within the general nonlinear iteration solution
they should be summarized (accumulated) into the full gluon
propagator and effectively correctly described by its structure in
the deep IR domain, exactly represented by its INP part (4.21).
The second step is to assign a mathematical meaning to the
integrals, where such kind of severe IR singularities will
explicitly appear, i.e., to define them correctly in the IR region
\cite{10,17,18}. Just this violent IR behavior makes QCD as a
whole an IR unstable theory, and therefore it may have no IR
stable fixed point \cite{1}. This means that QCD itself might be a
confining theory without involving some extra degrees of freedom
\cite{12,20,21,22,23,24,25,26}. In part II we will show that this
is so, indeed.

It is instructive to discuss the principal difference between QED
and QCD from the point of view of the structures of the full
photon and gluon propagators, respectively. In section II it has
been explained that the mass gap cannot occur in QED because the
photon is a physical state. As a result, the full photon
propagator can have only the PT-like IR singularity, i.e.,
$1/q^2$, see Eq. (2.13). This is in agreement with the cluster
property of the Wightman functions, that is, correlation functions
of observables in QED \cite{27}. It forbids a more singular
behavior of the full photon propagator in the IR than the behavior
of its free photon counterpart. In QCD, due to color confinement
the gluon is not a physical state. From our investigation then it
follows that the mass gap can be realized, and the general
nonlinear iteration solution for the full gluon propagator (4.20)
becomes inevitably severely IR singular. This is in agreement with
the so-called Strocchi theorem \cite{28}, which validates such
singular structure of the full gluon propagator in the IR. So the
existence of a mass gap is a primary reason for a possible
violation of the cluster property of the Wightman functions in QCD
(to our best knowledge there is no rigorous proof of this property
in this theory, see discussion in Ref. \cite{5} as well). At the
fundamental quark-gluon level only these remarks about the
structure of the Wightman functions in QCD make sense before the
solution of the color confinement problem. Concluding, let us note
in advance that in the presence of a mass gap there also exists
the so-called massive solution. It becomes explicitly smooth at
small $q^2$ in the Landau gauge only (see appendix A).

The PT part of the full gluon propagator (4.23), which has the
power-type PT IR singularity only, remains undetermined. Its
functional structure is due to the unknown in general $\Pi(q^2;
D)$ term in Eq. (4.1). This is the price we have paid to fix
exactly the functional dependence of the INP part of the full
gluon propagator. What we know about the PT effective charge
$d^{PT}(q^2; \Delta^2)$ is that it cannot have the power-type IR
singularities; otherwise it remain arbitrary. Also, in the formal
PT $\Delta^2=0$ limit it survives, i.e., $d^{PT}(q^2; \Delta^2=0)
= d^{PT}(q^2)$. This can be shown directly by restoring the
explicit dependence on the mass gap instead the dependence on the
coefficient $c= (\Delta^2 / M^2)$, which appears in the
splitting/shifting procedure. According to this procedure, the PT
effective charge can be explicitly written down as follows:

\begin{equation}
d^{PT}(q^2; \Delta^2) = \sum_{k=0}^{\infty} \Bigl( {\Delta^2 \over
M^2} \Bigr)^k \sum_{m=0}^{\infty} A_{km}(q^2),
\end{equation}
where $A_{km}(q^2)$ are dimensionless functions, which cannot have
the power-type IR singularities; otherwise they remain arbitrary
(their dependence on the parameters are not shown, for
simplicity). The explicit presence of the mass gap (or
equivalently the above-mentioned coefficient $c$) in this
expansion just prevents the ghosts to cancel the longitudinal
component in the full gluon propagator (4.20). However, in the
formal PT $ \Delta^2=0$ limit the role of ghosts will be
automatically restored, as pointed out above in section III. At
the same time, it is worth emphasizing that due to the character
of the IR singularity the longitudinal component of the full gluon
propagator should be included into its PT part. That is the reason
why its INP part becomes automatically transversal, as emphasized
above. This also means that the PT part of the full gluon
propagator (4.23) contains the free PT gluon propagator, so that
we can put $d^{PT}(q^2) = 1 + d^{AF}(q^2)$, where $d^{AF}(q^2)$
should satisfy AF in the $q^2 \rightarrow \infty$ limit (see
appendix B).

Both terms in Eq. (4.20) are valid in the whole energy/momentum
range, i.e., they are not asymptotics. At the same time, we have
achieved the exact separation between the two terms responsible
for the NP (dominating in the IR ($q^2 \rightarrow 0$)) and the
nontrivial PT (dominating in the UV ($q^2 \rightarrow \infty$))
dynamics in the true QCD vacuum. The structure of this solution
clearly confirms our conclusion driven above that the deep IR
region interesting for confinement and other NP effects is
dominated by the presence of a mass gap. In the formal PT
$\Delta^2=0$ limit, the nontrivial PT dynamics is all that
matters. Let us note in advance that the separation is not only
exact but it is unique as well. There exists a special
regularization expansion for severe (i.e., NP) IR singularities,
while for the PT IR singularity, which is only one present in the
PT part of the full gluon propagator (4.23), such kind of
expansion does not exist \cite{17,19}. We came to the same
structure (4.20)-(4.23) but in a rather different way in Refs.
\cite{10,18,19}.

In summary, the separation between the INP and PT terms in the
full gluon propagator (4.20) is exact and unique. Despite the
explicit dependence on the mass gap the gluons remain massless.
Moreover, the general nonlinear iteration solution (4.20) is
inevitably severely singular in the IR limit ($q^2 \rightarrow
0$), and this does not depend on the special gauge choice.

\section{General discussion}

The mass gap $\Delta^2 \equiv \Delta^2(\lambda, \alpha, \xi, g^2)$
has not been introduced by hand. It is hidden in the skeleton loop
integrals, contributing to the full gluon self-energy. No
truncations/approximations/assumptions and no special gauge choice
are made for these integrals. An appropriate subtraction scheme
has been applied to make the existence of a mass gap perfectly
clear. It is dynamically generated mainly by the NL interaction of
massless gluon modes. The Lagrangian of QCD does not contain a
mass gap, while it explicitly appears in the gluon SD equation of
motion. This once more underlines the importance of the
investigation of the SD system of equations and the corresponding
ST identities (\cite{1,8,10,29} and references therein) for
understanding of the true structure of the QCD ground state.

From the relation (3.19) it follows that the mass gap shows up
explicitly when the gluon momentum goes to zero. Especially this
is clearly seen in the general nonlinear iteration solution
(4.20), where it indeed explicitly determines its IR structure.
There are no doubts that the mass gap plays a dominant role in the
dynamics of QCD at large distances. So the problem how to
"liberate/release" it from the QCD vacuum becomes vitally
important to understand correctly the mechanism of color
confinement. It turned out that for this aim we need to sacrifice
the role of ghosts at the initial stage, while restoring the
transversality of the full gluon propagator relevant for NP QCD at
the final stage.

In order to realize a mass gap (more precisely its regularized
version), we propose not to impose the transversality condition on
the full gluon self-energy, see Eq. (3.14), while always
preserving the ST identity for the full gluon propagator, see Eq.
(3.15). Such a self-consistent violation of color gauge
invariance/symmetry (SCVCGI/S) is completely NP effect, since in
the formal PT $\Delta^2=0$ limit this effect vanishes. The first
necessary condition for the SCVCGI/S is color confinement, due to
which the gluon is not a physical state. The second sufficient
condition for the SCVCGI/S is the explicit presence of the
skeleton tadpole term in the full gluon self-energy. In other
words, not to request the transversality of the gluon self-energy
is the first necessary condition to realize a mass gap. In this
case whether the tadpole term is explicitly present or not in the
gluon self-energy becomes not important. It plays only secondary
role, indicating that the above-mentioned transversality may be
violated at the initial stage, indeed. All this makes it possible
to establish the structure of the regularized full gluon
propagator and the corresponding gluon SD equation in the presence
of a mass gap. After the restoration of the transversality of the
full gluon propagator relevant for NP QCD in part II of our
investigation, the SCVCGI/S in QCD will have no direct physical
consequences. None of physical quantites/processes in low-energy
QCD will be affected by this proposal, i.e., the unitarity of
$S$-matrix in NP QCD will not suffer, as emphasized above.

In QED a mass gap is always in the "gauge prison". It cannot be
realized even temporarily, since the photon is a physical state.
However, in QCD a door of the "color gauge prison" can be opened
for a moment in order to realize a mass gap, because the gluon is
not a physical state. A key to this "door" is the constant
skeleton tadpole term, which explicitly violates the
transversality of the full gluon self-energy. On the other hand,
this "door" can be opened without a key (as any door) by not
imposing the transversality condition on the full gluon
self-energy. So in QED a mass gap cannot be "liberated/released"
from the vacuum, while photons and electrons can be
liberated/released from the vacuum in order to be physical states.
In QCD a mass gap can be "liberated/released" from the vacuum,
while gluons and quarks cannot be liberated/released from the
vacuum in order to be physical states. In other words, there is no
breakdown of $U(1)$ gauge symmetry in QED because the photon is a
physical state. At the same time, a temporary breakdown of $SU(3)$
color gauge symmetry in QCD is possible because the gluon is not a
physical state.

In summary, QCD as a theory of quark-gluon interactions may have a
mass gap $\Delta^2$, possibly realized in accordance with our
proposal. The dynamically generated mass is usually related to
breakdown of some symmetry (for example, the dynamically generated
quark mass is an evidence of chiral symmetry breakdown). Here a
mass gap is an evidence of the SCVCGI/S. In the presence of a mass
gap the coupling constant plays no role. Thus the SCVCGI/S is also
a direct evidence of the "dimensional transmutation", $g^2
\rightarrow \Delta^2(\lambda, \alpha, \xi, g^2)$ \cite{1,30,31},
which occurs whenever a massless theory acquires masses
dynamically. It is a general feature of spontaneous symmetry
breaking in field theories.

\section{Conclusions}

A self-consistent violation of $SU(3)$ color gauge
invariance/symmetry for the realization of a mass gap as it has
been described in this investigation (section III) and briefly
discussed in the previous section is our {\bf first} main result.

The structures of the regularized full gluon propagator and the
corresponding gluon SD equation in the presence of a mass gap have
been firmly established. This is our {\bf second} main result
(subsection C in section III).

The general nonlinear iteration solution (4.20) for the full gluon
propagator explicitly depends on the mass gap. However, it is
always severely singular in the $q^2 \rightarrow 0$ limit. So the
gluons remain massless, and this does not depend on the gauge
choice. This solution is our {\bf third} main result (section IV).

In the presence of a mass gap gluon may acquire an effective gluon
mass, depending on the gauge choice (the so-called massive
solution). This is our {\bf fourth} main result. Nevertheless, we
put it into appendix A, since its relation to the solution of the
color confinement problem is unclear, even after the
renormalization program is performed.

Thus, we have shown explicitly that in the presence of a mass gap
at least two independent and different types of formal solutions
for the regularized full gluon propagator exist (let us remind
that the gluon SD equation is highly NL one, so the number of
independent solutions is not fixed). No
truncations/approximations/assumptions and no special gauge choice
are made in order to show the existence of these general types of
solutions.

The nonlinear iteration solution (4.20) is interesting for
confinement. However, three important problems remain to solve:

{\bf A}. How to make this solution to be the purely transversal.

{\bf B}. To perform the renormalization program for the mass gap,
and to see whether the mass gap survives it or not.

{\bf C}. How to treat correctly severe IR singularities inevitably
present in this solution.

\vspace{3mm}

Let us now explain each of these points in more detail.

{\bf A}. As we already know, in the presence of a mass gap the
ghosts alone cannot provide the cancellation of the longitudinal
component of the full gluon propagator. So a universal method of
the restoration of the transversality of the full gluon propagator
relevant for NP QCD should be formulated. This will make it
possible to eliminate the explicit dependence on the gauge-fixing
parameter from all the corresponding expressions in NP QCD. It is
worth emphasizing in advance that due to the above-mentioned
universal method, the PT part of the general nonlinear iteration
solution (4.23), which remains undetermined, will be of no
importance for us. Only its INP counterpart (4.21)-(4.22), which
functional dependence is exactly determined up to the
above-mentioned residues, will matter.

{\bf B}. As repeatedly mentioned above, in this paper the mass gap
has been only regularized, i.e., $\Delta^2 \equiv
\Delta^2(\lambda, \alpha, \xi, g^2)$. To perform the
renormalization program means to remove the dependence on the
above shown parameters in a self-consistent way. We should prove
that the product $\Delta^2_{JW} = Z(\lambda, \alpha, \xi, g^2)
\Delta^2(\lambda, \alpha, \xi, g^2)$ exists in the $\lambda
\rightarrow \infty$ and $\alpha \rightarrow 0$ limits. The mass
gap's renormalization constant $Z(\lambda, \alpha, \xi, g^2)$
should appear naturally, i.e., it should not be introduced by hand
in order not to compromise the general renormalizability of QCD.
In other words, the renormalized mass gap should not depend on the
gauge-fixing parameter, should be finite, positive definite, etc.
Only after performing this program, we can assign to the
Jaffe-Witten (JW) mass gap $\Delta^2_{JW}$ a physical meaning to
be responsible for the NP dynamics in QCD.

{\bf C}. This solution is characterized by the explicit presence
of severe IR singularities $(q^2)^{-2-k}, \ k=0,1,2,3...$, which,
in principle, are independent from each other. The only method to
treat them in a self-consistent way is the DRM \cite{11},
correctly implemented into the DT \cite{16}, as emphasized above.

All these problems will be addressed and solved in our subsequent
paper (part II of our investigation).

Concluding, a few preliminary remarks are in order. We have
already emphasized that the most appropriate place where the mass
gap may appear is the system of the SD equations (complemented by
the corresponding ST identities) for the QCD Green's functions in
momentum space. The gauge dependence of the gluon Green's function
(4.20) in this space is in general twofold: the explicit
dependence on the gauge-fixing parameter via the longitudinal
component and the implicit gauge dependence in the Lorentz
structure, affiliated with its transversal component. Within our
approach the former will be solved if we will be able to formulate
a general method how to restore the transversality of the full
gluon propagator relevant for NP QCD (the problem {\bf A}). The
latter one will be solved if we will be able to find how to
renormalize the mass gap in a gauge-invariant way, i.e., the mass
gap should survive the renormalization program (the problem {\bf
B}). Solving the above-mentioned problems, we will achieve an
explicit gauge independence at the fundamental quark-gluon level,
while it is too earlier to discuss the gauge independence of
$S$-matrix elements in NP QCD within our approach. Working always
in momentum space, we avoid thus the problem of gauge ambiguity
(uncertainty) \cite{32}, which has been discovered in much more
complicated functional space (see Ref. \cite{33} as well).

\begin{acknowledgments}

Support by HAS-JINR grant (P. Levai) is to be acknowledged. The
author is grateful to P. Forgacs, J. Nyiri, and especially to H.
Toki for useful discussions, remarks and help.

\end{acknowledgments}

\appendix

\section{Massive solution}

One of the direct consequences of the explicit presence of a mass
gap in the full gluon propagator is that the gluon may indeed
acquire an effective mass. From Eq. (3.19) it follows that

\begin{equation}
{ 1 \over q^2} d(q^2) = {1 \over q^2 + q^2 \Pi(q^2; \xi) +
\Delta^2(\lambda, \xi)},
\end{equation}
where instead of the dependence on $D$ the dependence on $\xi$ is
explicitly shown (let us remind that as in the case of the
iteration solution, we replace $\Pi^s(q^2; \xi) \rightarrow
\Pi(q^2; \xi)$). The full gluon propagator (3.16) may have a
pole-type solution at the finite point if and only if the
denominator in Eq. (A1) has a zero at this point $q^2 = - m^2_g$
(Euclidean signature), i.e.,

\begin{equation}
- m^2_g  - m^2_g \Pi(-m^2_g; \xi) + \Delta^2(\lambda, \xi)=0,
\end{equation}
where $m^2_g \equiv m^2_g(\lambda, \xi)$ is an effective gluon
mass, and the previous equation is a transcendental equation for
its determination. Evidently, the number of the solutions of this
equation is not fixed, $a \ priori$. Excluding the mass gap, one
obtains that the denominator in the full gluon propagator becomes

\begin{equation}
q^2 + q^2 \Pi(q^2; \xi) + \Delta^2(\lambda, \xi) = q^2 + m^2_g +
q^2 \Pi(q^2; \xi) + m^2_g \Pi(-m^2_g; \xi).
\end{equation}

Let us now expand $\Pi(q^2; \xi)$ in a Taylor series near $m^2_g$:

\begin{equation}
\Pi(q^2; \xi) = \Pi(-m^2_g; \xi) + (q^2 + m^2_g) \Pi'(-m^2_g; \xi)
+ O \Bigl( (q^2 + m^2_g)^2 \Bigr).
\end{equation}
Substituting this expansion into the previous relation and after
doing some tedious algebra, one obtains

\begin{equation}
q^2 + m^2_g + q^2 \Pi(q^2; \xi) + m^2_g \Pi(-m^2_g; \xi)= (q^2 +
m^2_g)[1 +  \Pi(-m^2_g; \xi) - m^2_g \Pi'(-m^2_g; \xi)] [ 1 +
\Pi^R(q^2; \xi)],
\end{equation}
where $\Pi^R(q^2; \xi)= 0$ at $q^2=-m^2_g$; otherwise it remains
arbitrary. Thus the full gluon propagator (3.16) now looks

\begin{equation}
D_{\mu\nu}(q; m^2_g) = i T_{\mu\nu}(q) {Z_3 \over (q^2 + m^2_g) [
1 + \Pi^R(q^2; m^2_g)]} + i \xi L_{\mu\nu}(q) {1 \over q^2},
\end{equation}
where, for future purpose, in the invariant function $\Pi^R(q^2;
m^2_g)$ instead of $\xi$ we introduced the dependence on the gluon
effective mass squared $m_g^2$ which depends on $\xi$ itself. The
gluon renormalization constant is

\begin{equation}
Z_3 = { 1 \over 1 + \Pi(-m^2_g; \xi) - m^2_g \Pi'(-m^2_g; \xi)}.
\end{equation}
In the formal PT limit $\Delta^2(\lambda, \xi) =0$, an effective
gluon mass is also zero, $m_g^2(\lambda, \xi) =0$, as it follows
from Eq. (A2). So an effective gluon mass is the NP effect. At the
same time, it cannot be interpreted as the "physical" gluon mass,
since it remains explicitly gauge-dependent quantity (at least at
this stage). In other words, we were unable to renormalize it
along with the gluon propagator (A6). In the formal PT
$\Delta^2(\lambda, \xi) =m_g^2(\lambda, \xi) =0$ limit the gluon
renormalization constant (A7) becomes the standard one
\cite{2,8,9}, namely

\begin{equation}
Z_3^{PT} = {1 \over 1 + \Pi(0; \xi)},
\end{equation}
and the role of ghosts will be automatically restored, as pointed
out above in section III.

Concluding, it is interesting to note that Eq. (A2) has a second
solution in the formal PT $\Delta^2(\lambda, \xi) =0$ limit. In
this case an effective gluon mass remains finite, but $1 +
\Pi(-m^2_g; \xi) =0$. So a scale responsible for the NP dynamics
is not determined by an effective gluon mass itself, but by this
condition. Its interpretation from the physical point of view is
not clear. Evidently, the massive solution (A6) is completely
independent from the general nonlinear iteration solution (4.20),
and it is difficult to relate it to confinement. However, its
existence shows the general possibility for a vector particles to
acquire masses dynamically, i.e., without so-called Higgs
mechanism (which in its turn requires the existence of not yet
discovered Higgs particle). The above-mentioned possibility is due
only to the internal dynamics and symmetries of the corresponding
gauge theory.

\section{Remarks on Asymptotic Freedom}

As mentioned above, our general solution for the "running"
effective charge (3.19) should satisfy AF at large $q^2$. We
already know that in the $q^2 \rightarrow \infty$ limit the
subtracted gluon self-energy $\Pi^s(q^2; D)$ can be only
logarithmically divergent at any $D$. So neglecting the mass gap
term in the relation (3.19), to leading order one obtains

\begin{equation}
g^2(q^2; \Lambda^2) = { g^2(\lambda) \over 1 + b g^2(\lambda) \ln
(q^2 / \Lambda^2)},
\end{equation}
where $\Lambda^2$ is the UV cutoff squared, and $b>0$ is the
standard color group factor \cite{1,2}. We introduce into the
numerator $g^2(\lambda)$, so that when it formally zero (no
interaction at all) then the full gluon propagator should be
reduced to its free PT counterpart. For convenience, we also
denote $d(q^2)$ as $g^2(q^2; \Lambda^2)$, i.e., put $d(q^2) \equiv
g^2(q^2; \Lambda^2)$. This expression represents the summation of
the so-called main PT logarithms. However, nothing should depend
on $\Lambda$ (and hence on $\lambda$) when they go to infinity in
order to recover the finite effective charge in this limit. To
show explicitly that this finite limit exists, let us rewrite the
previous expression in the symmetric form \cite{34}

\begin{equation}
{ g^2(\lambda_1) \over 1 + b g^2(\lambda_1) \ln (q^2 /
\Lambda^2_1)} = { g^2(\lambda_2) \over 1 + b g^2(\lambda_2) \ln
(q^2 / \Lambda^2_2)},
\end{equation}
since $g^2(q^2; \Lambda^2_1)=g^2(q^2; \Lambda^2_2)$, i.e., nothing
should depend on how we denote the UV cutoff, indeed. In the
$\Lambda_{1,2} \rightarrow \infty$ (and hence $\lambda_{1,2}
\rightarrow \infty$) limits neglecting the dependence on $\ln
q^2$, from the relation (B2) one obtains

\begin{equation}
\ln \Lambda_2 - {1 \over 2 b g^2(\lambda_2)} = \ln \Lambda_1 - {1
\over 2 b g^2(\lambda_1)},
\end{equation}
and this relation becomes more and more exact with all the UV
cutoffs becoming bigger and bigger (and thus the suppression of
$\ln q^2$ becoming more and more justified). Evidently, this
relation is equivalent to

\begin{equation}
\Lambda_2 \exp \Bigl( - {1 \over 2 b g^2(\lambda_2)} \Bigr) =
\Lambda_1 \exp \Bigl( - {1 \over 2 b g^2(\lambda_1)} \Bigr).
\end{equation}
Thus there exists indeed the limit

\begin{equation}
\lim_{(\Lambda, \lambda) \rightarrow \infty} \Lambda \exp \Bigl( -
{1 \over 2 b g^2(\lambda)} \Bigr) = \Lambda_{QCD}
\end{equation}
at which it is finite and does not depend on the UV cutoff or the
renormalization point (evidently, not losing generality, we can
estimate that $2 b g^2(\lambda) \sim 1 / \ln \lambda$ in the
$\lambda \rightarrow \infty$ limit). This finite limit is nothing
but $\Lambda_{QCD} = \Lambda_{PT}$, which governs the nontrivial
dynamics of PT QCD in asymptotic regime at large $q^2$. Thus,
using the limit (B5), we can rewrite the initial expression (B1)
in terms of the finite quantities. It reproduces the well known AF
behavior of the effective charge in QCD ar large $q^2$, namely

\begin{equation}
\alpha_s(q^2) = {1  \over b \ln (q^2 / \Lambda^2_{QCD})},
\end{equation}
where the standard notation $\alpha_s(q^2) = g^2(q^2) / 4 \pi$ has
been used. It can be generalized like Eq. (B1) as follows:

\begin{equation}
\alpha_s(q^2) = {\alpha_s \over 1 + b \alpha_s \ln (q^2 /
\Lambda^2_{QCD})},
\end{equation}
where $\alpha_s = g^2 / 4 \pi$ is the fine-structure constant of
strong interactions, calculated at some fixed scale, for example
at $Z$ boson mass.  At a very large $q^2$ one recovers the
previous expression.

Concluding, within our approach we have shown explicitly the AF
behavior of QCD at short distances ($q^2 \rightarrow \infty$) not
using the renormalization group equations \cite{1,2,8,9,34}. There
is no relation between the mass gap (even renormalized) and the
asymptotic scale parameter, at least to leading order.

\section{Remarks on overlapping divergences}

In order to unravel overlapping UV divergences problem in QCD, in
each of the standard SD equations the necessary number of the
differentiation with respect to the external momentum should be
done first (in order to lower divergences). Then the point-like
vertices, which are present in the corresponding skeleton loop
integrals should be replaced by their full counterparts via the
corresponding integral equations. Finally, one obtains the
corresponding SD equations which are much more complicated than
the previous (standards) ones, containing different scattering
amplitudes. These skeleton expansions are, however, free of the
above-mentioned overlapping divergences. Of course, the real
procedure (\cite{16} and references therein) is much more tedious
than briefly described above.

However, even at this level it is clear that by taking derivatives
with respect to the external momentum $q$ in the gluon SD equation
(3.17), which is convenient to rewrite as follows:

\begin{equation}
D^{-1}(q) = D^{-1}_0(q) - q^2 \Pi(q^2; D) - \Delta^2(\lambda; D),
\end{equation}
the main initial information on the mass gap will be totally lost
(we omit the tensor indices, for simplicity). Just the mass gap
which appears first in this equation is the main object we were
worried about to demonstrate explicitly its crucial role within
our approach. Whether the above-mentioned information will be
somehow restored or not at the later stages of the renormalization
program is not clear at all. Thus in order to remove overlapping
UV divergences ("the water") from the SD equations and skeleton
expansions, we are in danger to completely lose the information on
the dynamical source of the mass gap ("the baby") within our
approach. In order to avoid this danger and to be guaranteed that
no dynamical information are lost, we are using the standard gluon
SD equation (3.17) in the presence of a mass gap. The existence of
any kind of UV divergences (overlapping and usual (overall)) in
the skeleton expansions will not cause any problems in order to
detect the mass gap responsible for the IR structure of the true
QCD vacuum. As emphasized above, the problem of convergence of the
regularized skeleton loop series which appear in Eq. (4.20) is
completely irrelevant in the context of the present investigation.
Anyway, we keep any kind of UV divergences under control within
our method, since we are working with the regularized quantities.
At the same time, the existence of a mass gap responsible for the
IR structure of the full gluon propagator does not depend on
whether overlapping divergences are present or not in the SD
equations and corresponding skeleton expansions. As argued above,
the existence of a mass gap is only due to the SCVCGI/S. All this
is the main reason why our starting point is the standard gluon SD
equation (3.1) for the unrenormalized (but necessarily
regularized) Green's functions (this also simplifies notations).

\end{document}